\documentclass[journal=jacsat,manuscript=article]{achemso}
\usepackage{xr}
\makeatletter
\newcommand*{\addFileDependency}[1]{
  \typeout{(#1)}
  \@addtofilelist{#1}
  \IfFileExists{#1}{}{\typeout{No file #1.}}
}
\makeatother

\newcommand*{\myexternaldocument}[1]{
    \externaldocument{#1}
    \addFileDependency{#1.tex}
    \addFileDependency{#1.aux}
}
\newcommand{\zcite}[1]{\scalebox{1.5}[1.5]{\raisebox{-0.84ex}{\cite{#1}}}} 

\myexternaldocument{Supportinginformation}

\usepackage[version=3]{mhchem} 
\usepackage[T1]{fontenc} 
\usepackage{gensymb}

\author{Shili Yan}
\affiliation[BAQIS]
{Beijing Academy of Quantum Information Sciences, 100193 Beijing, China}

\author{Yi Luo}
\affiliation[PKU]
{Beijing Key Laboratory of Quantum Devices and School of Electronics, Peking University, Beijing 100871, China}
\alsoaffiliation[ICMMP]
{Institute of Condensed Matter and Material Physics, School of Physics, Peking University, Beijing 100871, China}

\author{Haitian Su}
\affiliation[PKU]
{Beijing Key Laboratory of Quantum Devices and School of Electronics, Peking University, Beijing 100871, China}
\alsoaffiliation[ICMMP]
{Institute of Condensed Matter and Material Physics, School of Physics, Peking University, Beijing 100871, China}

\author{Han Gao}
\affiliation[BAQIS]
{Beijing Academy of Quantum Information Sciences, 100193 Beijing, China}

\author{Xingjun Wu}
\affiliation[BAQIS]
{Beijing Academy of Quantum Information Sciences, 100193 Beijing, China}

\author{Dong Pan}
\email{pandong@semi.ac.cn}
\affiliation[CASIOS]
{State Key Laboratory of Semiconductor Physics and Chip Technologies, Institute of Semiconductors,Chinese Academy of Sciences, P.O. Box 912, Beijing 100083, China}

\author{Jianhua Zhao}
\affiliation[CASIOS]
{State Key Laboratory of Semiconductor Physics and Chip Technologies, Institute of Semiconductors,Chinese Academy of Sciences, P.O. Box 912, Beijing 100083, China}

\author{Ji-Yin Wang}
\email{wang_jy@baqis.ac.cn}
\affiliation[BAQIS]
{Beijing Academy of Quantum Information Sciences, 100193 Beijing, China}

\author{Hongqi Xu}
\email{hqxu@pku.edu.cn}
\affiliation[BAQIS]
{Beijing Academy of Quantum Information Sciences, 100193 Beijing, China}
\alsoaffiliation[PKU]
{Beijing Key Laboratory of Quantum Devices and School of Electronics, Peking University, Beijing 100871, China}

\title{Gate Tunable Josephson Diode Effect in Josephson Junctions made from InAs Nanosheets }

\keywords{Josephson Diode Effect, Rashba Spin-orbit Interaction, InAs Nanosheets\LaTeX}

\begin{document}
\newpage

\maketitle
\normalsize
\begin{abstract}
We report the observation of Josephson diode effect (JDE) in hybrid devices made from semiconductor InAs nanosheets and superconductor Al contacts. By applying an in-plane magnetic field ($B_{\mathrm{xy}}$), we detect non-reciprocal superconducting switching current as well as non-reciprocal superconducting retrapping current. The strength of the JDE depends on the angle between the in-plane magnetic field and the bias current ($I_{\mathrm{b}}$), reaching its maximum when $B_{\mathrm{xy}} \perp I_{\mathrm{b}}$ and dropping to nearly zero when $B_{\mathrm{xy}}\parallel  I_{\mathrm{b}}$. Additionally, the diode efficiency is tunable via an electrostatic gate with a complete suppression at certain gate voltages. Our findings indicate that the observed JDE in InAs nanosheet-based Josephson junctions most likely arises from the Rashba spin-orbit interaction (SOI) in the nanosheets. Such gate-tunable JDE in Josephson junctions made from semiconductor materials with SOI is useful not only for constructing advanced superconducting electronics but also for detecting novel superconducting states.


\end{abstract}
\newpage

\section{Introduction}
\par Superconducting diode effect (SDE) refers to a non-reciprocal behavior of the superconducting critical current with respect to the current directions. Analogous to the role of traditional p-n junction diodes in semiconductor electronics, SDE is regarded as a key functional element for superconducting circuits.\cite{ linder2015superconducting, braginski2019superconductor, wendin2017quantum,nadeem2023superconducting} Such a non-reciprocal superconducting effect in Josephson junction devices, specified as Josephson diode effect (JDE), can also be used to identify the presence of topological superconductivity.\cite{legg2023parity}  SDE/JDE has therefore garnered significant attention from both theoretical and experimental perspectives in recent years.\cite{ando2020observation,legg2023parity,he2022phenomenological,daido2022intrinsic,yuan2022supercurrent, davydova2022universal, legg2022superconducting,ilic2022theory,souto2022josephson,PhysRevResearch5033131,pal2022josephson,baumgartner2022supercurrent,bauriedl2022supercurrent,lin2022zero,turini2022josephson,mazur2022gate,golod2022demonstration,gutfreund2023direct, ciaccia2023gate, hou2023ubiquitous,banerjee2023phase,margineda2023sign,costa2023sign,sundaresh2023diamagnetic,valentini2024parity,su2024microwave,lotfizadeh2024superconducting,wu2025tunable} In general, the occurrence of SDE/JDE requires the simultaneous breaking of time-reversal and inversion symmetries.\cite{edelstein1996ginzburg} This can arise from intrinsic mechanisms, such as finite-momentum pairing,\cite{he2022phenomenological,daido2022intrinsic,yuan2022supercurrent, davydova2022universal, legg2022superconducting,ilic2022theory} or extrinsic factors like device geometry or magnetic flux tuning in superconducting quantum interference devices (SQUIDs).\cite{goldman1967meissner, hou2023ubiquitous,golod2022demonstration, ciaccia2023gate,valentini2024parity,souto2022josephson}  Among these, finite-momentum pairing is of particular interest, as it can be induced by the interplay between spin-orbit interaction (SOI) and Zeeman energy.\cite{ yuan2022supercurrent, ilic2022theory,pal2022josephson, turini2022josephson, mazur2022gate, lotfizadeh2024superconducting} In this case, when a magnetic field is applied parallel to the effective SOI field, the system enters a novel superconducting state with finite-momentum pairing at the Fermi surface. In a Josephson junction, an anomalous superconducting phase $\varphi_{\mathrm{0}}$ accumulates in this configuration when Cooper pairs tunnel through the junction. 
\cite{szombati2016josephson,ke2019josephson} Combined with high interface transparency in the junction, this anomalous phase can give rise to JDE.\cite{ilic2022theory, zhang2022evidence, costa2023microscopic} 
\par Semiconductor nanostructures with strong SOI are particularly promising platforms for studying the JDE due to their high gate tunability and potential connection to topological superconductivity.\cite{pientka2017topological,dartiailh2021phase} Pioneering research has been carried out in Josephson junctions made of semiconductor nanostructures with strong SOI, such as InAs and InSb nanowires or two-dimensional electron gases\cite{zhang2022evidence,turini2022josephson,mazur2022gate,lotfizadeh2024superconducting,shin2024electric}. The studies of Sn-InSb nanowires and Nb-InSb nanoflags have confirmed the presence of the JDE but demonstrated limited gate tunability.\cite{ zhang2022evidence,turini2022josephson} In contrast, Josephson junctions made from InSb nanowires and superconductor Al leads have exhibited a substantial and complex dependence of the JDE on gate voltages.\cite{mazur2022gate} In planar Josephson junctions made from epitaxial Al-InAs heterostructures, one research work demonstrated that the strength of the JDE varies with gate voltage, which implies the presence of gate-tunable Rashba SOI.\cite{lotfizadeh2024superconducting} Another study reported a weak dependence of the JDE on gate voltage under an in-plane magnetic field, when the magnetic field strength remained below the threshold required for polarity reversal.\cite{shin2024electric} Despite these advances, electrostatic control of the JDE remains an active area of research, essential for both understanding exotic quantum states in hybrid superconductor-semiconductor systems and enabling future quantum device applications.
\par Here we demonstrate an intrinsic JDE in planar Josephson junctions made from InAs nanosheets. The JDE of the InAs nanosheet Josephson junctions shows a strong dependence on the magnetic field orientation as well as the magnitude of the field. A full rotation of the in-plane magnetic field orientation reveals that the diode efficiency is maximized when the magnetic field is perpendicular to the current direction. Furthermore, the diode efficiency is found to decrease almost monotonically as the back gate voltage decreases and reaches nearly zero at a specific gate voltage. These behaviors are consistent with finite-momentum pairing induced by the interplay of the Rashba SOI and the Zeeman effect. Our study demonstrates that the systems with SOI, Zeeman effect and induced superconductivity are highly interesting for the application in advanced superconducting electronics and for the exploration of novel superconducting states.\cite{ braginski2019superconductor, linder2015superconducting, wendin2017quantum}
\section{Results and discussion}
\subsection{JDE in InAs Nanosheet-Al Josephson Junctions}
\par
\begin{figure}[bt] 
\centering
\includegraphics[width=0.85\linewidth]{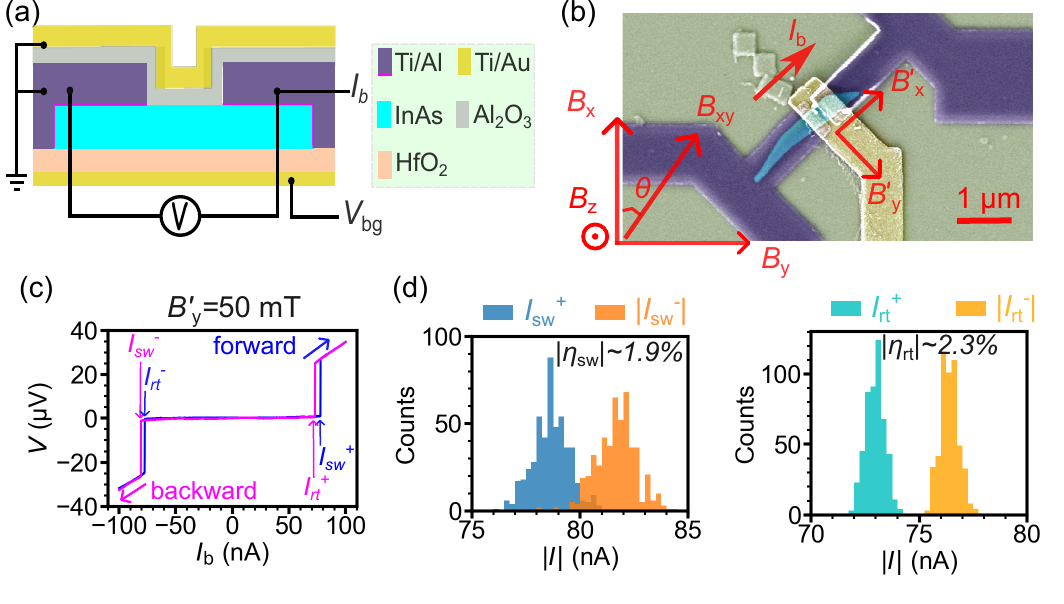}
\caption{\textbf{The Josephson diode effect (JDE) in InAs nanosheet-Al Josephson junctions.} \textbf{(a)} Schematic side view of the InAs nanosheet-Al Josephson junction device and the measurement circuit. The InAs nanosheet is colored in cyan. The source and drain electrodes colored in purple are made of 5/50\,nm Ti/Al. The back gate and top gate colored in gold are made of Ti/Au. The dielectric layers are colored gray and pink for $\mathrm{Al_2O_3}$ and $\mathrm{HfO_2}$, respectively. A quasi-four-terminal measurement setup is employed, where a current $I_{\mathrm{b}}$ is applied and the voltage $V$ across the junction is measured. The top gate is grounded throughout all measurements. \textbf{(b)} False-colored scanning electron microscope (SEM) image of device A. The red arrow,  labeled $I_{\mathrm{b}}$, indicates the current flow direction for positive bias current. $B_{\mathrm{x}}$, $B_{\mathrm{y}}$ and $B_{\mathrm{z}}$ are the axes for the coordinate of the vector magnet, and $B_{\mathrm{xy}}$ is the applied in-plane magnetic field. $B_{\mathrm{x}}^{'}$ and $B_{\mathrm{y}}^{'}$ correspond to a rotated coordinate system in which $B_{\mathrm{x}}^{'}$ is parallel to the current flow direction. \textbf{(c)} Representative $V$-$I$ curves illustrating the JDE at $B_{\mathrm{y}}^{'}=\rm{50}\,mT$ and $V_{\mathrm{bg}}=0$. Blue and pink curves correspond to forward and backward current sweeps, respectively. The switching currents ($I_{\mathrm{sw}}^{+}$ and $I_{\mathrm{sw}}^{-}$) as well as retrapping currents ($I_{\mathrm{rt}}^{+}$ and $I_{\mathrm{rt}}^{-}$) are marked. \textbf{(d)} Histograms of switching current (left panel) and retrapping current (right panel) over N=500 measurements at $B_{\mathrm{y}}^{'}=50\,\rm{mT}$ and $V_{\mathrm{bg}}=0$. The current ramp rate is $\sim$4.5\,nA/s.} 
\label{figure:1}
\end{figure}

 The  Josephson junction devices were fabricated from InAs nanosheets grown via molecular beam epitaxy (MBE).\cite{pan2019dimension} Typically, the nanosheets have a thickness of 15--30 nm with a wurtzite crystal structure. The mobility of the InAs nanosheets is approximately 6500-8000 cm$^2$/V$\cdot$s and the carrier density can be tuned from complete depletion up to $\sim$4 $\times$ 10$^{12}$ cm$^{-2}$
 (see Refs.\zcite{pan2019dimension} and \zcite{yan2023supercurrent}). The nanosheets were mechanically transferred onto pre-patterned local back gates, covered with a 15-nm-thick $\mathrm{{HfO_\mathrm{2}}}$ dielectric layer grown by atomic layer deposition (ALD), on a SiO$_2$/Si substrate. The contact areas of the nanosheets were defined by electron-beam lithography. Then, the samples were etched in a diluted $\mathrm{(NH_4)_2S_x}$ solution to remove the native oxide layer, followed by a deposition of 5 nm Ti using electron-beam evaporation and 50 nm Al using thermal evaporation. Top gates were fabricated with a deposition of 5/15 nm Ti/Au after growing 15 nm $\mathrm{Al_2O_3}$ with ALD. Figure \ref{figure:1}a displays a schematic side view of the fabricated devices together with the measurement circuit. Figure \ref{figure:1}b shows a false-colored scanning electron microscope (SEM) image of device A. For this device, the junction has a gap of $\sim$$100\,\mathrm{nm}$ between the superconducting electrodes and the InAs nanosheet at the junction is approximately $ 300\,\mathrm{nm}$ wide. Electrical transport measurements were performed on the device in a $^3\mathrm{He/}^4\mathrm{He}$ dilution refrigerator equipped with a vector magnet. In Figure \ref{figure:1}b, the axes $B_\mathrm{x}$, $B_\mathrm{y}$, and $B_\mathrm{z}$ indicate the coordinate of the vector magnet, while the axes $B_{\mathrm{x}}^{'}$ and $B_{\mathrm{y}}^{'}$ define a new coordinate with respect to the current direction $I_\mathrm{b}$. $B_\mathrm{xy}$ is an in-plane magnetic field with an angle $\theta$ with respect to $B_\mathrm{x}$. Two comparable devices (A and B) were studied, with all the results presented in the main article obtained from device A, while the results from device B are provided in the Supporting Information (Additional Data Section). Top-gates were grounded throughout this work.
\par Figure \ref{figure:1}c presents an example of the measured JDE at $B_{\mathrm{y}}^{'}=50\,\mathrm{mT}$. The voltage $V$ across the junction of the hybrid device is measured in a four-terminal measurement circuit setup while a bias current $I_{\mathrm{b}}$ is applied (see Figure \ref{figure:1}a). The blue and pink curves represent the results for forward and backward current sweeping directions, respectively. A characteristic hysteretic behavior is evident in both forward and backward sweeps, i.e. $I_{\mathrm{sw}}^{+} \neq I_{\mathrm{rt}}^{+} $ and $I_{\mathrm{sw}}^{-} \neq I_{\mathrm{rt}}^{-} $, which can be attributed to phase instability in the junction and/or heating effects.\cite{tinkham2004introduction,tinkham2003hysteretic,courtois2008origin,yan2023supercurrent,yan2023supercurrent} By comparing with previous studies,\cite{PhysRevLett.130.087002,PhysRevLett.109.050601} the switching mechanism of the device is likely to be phase diffusion. Aside from the hysteresis, the JDE is observed as a finite difference between the absolute values of forward sweep switching current $I_{\mathrm{sw}}^{+}\sim 77.7\,\mathrm{nA}$ and backward sweep switching current |$I_{\mathrm{sw}}^{-}|\sim 80.7\,\mathrm{nA}$. Additionally, there exists a finite difference between the absolute values of the retrapping current of forward and backward sweep $|I_{\mathrm{rt}}^{-}|\sim 77.3\,\mathrm{nA}$ and $I_{\mathrm{rt}}^{+}\sim 73.3\,\mathrm{nA}$, further confirming the presence of the JDE.\cite{lotfizadeh2024superconducting,su2024microwave} Since the switching current in a Josephson junction is typically stochastic\cite{doh2005tunable,lotfizadeh2024superconducting,su2024microwave}, repeated measurements are beneficial for an accurate analysis of the JDE. Figure \ref{figure:1}d presents histograms of switching current distribution, obtained from N=500 forward sweeping events followed by N=500 backward sweeping events under the same magnetic field conditions as in Figure \ref{figure:1}c. In the left histogram, there is a clear separation between $I_{\mathrm{sw}}^{+}$ and $|I_{\mathrm{sw}}^{-}|$ bunches. An averaged forward switching current, as well as backward switching current, is calculated from the distributions. The JDE efficiency of the switching current is then calculated, by using the averaged values, as $\eta_{\mathrm{sw}} = \Delta I_{\mathrm{sw}} /(|\!I_{\mathrm{sw}}^{-}\!| + I_{\mathrm{sw}}^{+}  )$, where $\Delta I_{\mathrm{sw}}=(|\!I_{\mathrm{sw}}^{-}\!| - I_{\mathrm{sw}}^{+})$. Similarly, the histograms for the retrapping current are shown in the right panel of Figure \ref{figure:1}d, and the diode efficiency of retrapping current is calculated as $\eta_{\mathrm{rt}} = \Delta I_{\mathrm{rt}} /(|\!I_{\mathrm{rt}}^{-}\!| + I_{\mathrm{rt}}^{+}  )$, with $\Delta I_{\mathrm{rt}}=(|\!I_{\mathrm{rt}}^{-}\!| - I_{\mathrm{rt}}^{+})$. Here, we obtain $\eta_{\mathrm{sw}}=1.9\%$ and $\eta_{\mathrm{rt}}=2.3\%$. These values are consistent with previous reports of the JDE in Josephson junctions made from low-dimensional semiconductors with strong SOI.\cite{turini2022josephson,mazur2022gate,lotfizadeh2024superconducting} Note that the variation of switching current is larger than that of retrapping current, potentially due to suppression of the stochastic behavior of retrapping current by heating effects.\cite{lotfizadeh2024superconducting,su2024microwave} The data presented for 
 $I_{\mathrm{sw}}^{+} $,  $I_{\mathrm{sw}}^{-}$, $I_{\mathrm{rt}}^{+}$ and $I_{\mathrm{rt}}^{+}$ in the rest of this article are averaged over repeated measurements similarly as done in Figure \ref{figure:1}d unless stated otherwise. 
\subsection{Magnetic Field Dependence}
\par
\begin{figure}[bt] 
\centering
\includegraphics[width=0.98\linewidth]{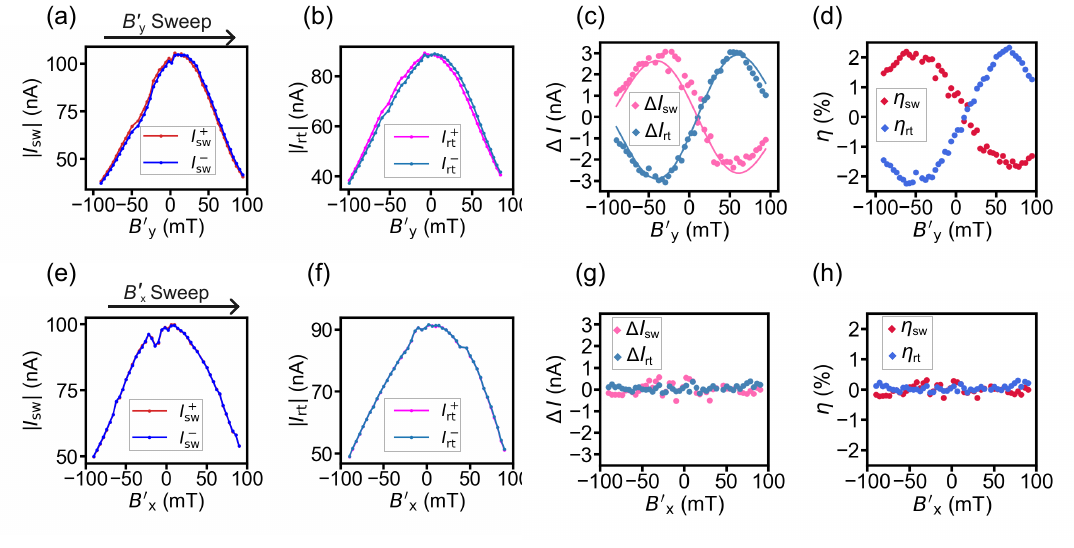}
\caption{\textbf{Dependence of the JDE on in-plane magnetic field strength.} 
\textbf{(a)} Absolute values of switching currents $I_{\mathrm{sw}}^{+}$ and $|I_{\mathrm{sw}}^{-}|$ as a function of $B_{\mathrm{y}}^{'}$. 
\textbf{(b)} Absolute values of retrapping currents $I_{\mathrm{rt}}^{+}$ and $|I_{\mathrm{rt}}^{-}|$ as a function of $B_{\mathrm{y}}^{'}$. 
\textbf{(c)} Differences of switching current ($\Delta I_{\mathrm{sw}}=(|I_{\mathrm{sw}}^{-}|-I_{\mathrm{sw}}^{+})$, pink dots) and retrapping current ($\Delta I_{\mathrm{rt}}=|I_{\mathrm{rt}}^{-}|-I_{\mathrm{rt}}^{+}$, blue dots) as a function of $B_{\mathrm{y}}^{'}$. The solid lines are fitting curves. 
\textbf{(d)} Diode efficiencies, $\eta_{\mathrm{sw}} = (|I_{\mathrm{sw}}^{-}|-I_{\mathrm{sw}}^{+})/(I_{\mathrm{sw}}^{+} + |I_{\mathrm{sw}}^{-}|)$ and $\eta_{\mathrm{rt}} = (|I_{\mathrm{rt}}^{-}| - I_{\mathrm{rt}}^{+})/(I_{\mathrm{rt}}^{+} + |I_{\mathrm{rt}}^{-}|)$, as a function of $B_{\mathrm{y}}^{'}$. 
Here note that the signs of $\Delta I_{\mathrm{sw}}$ and $\eta_{\mathrm{sw}}$ are inverted in the plot to facilitate clear visualization of quantities $\Delta I_{\mathrm{sw}}$ and $\Delta I_{\mathrm{rt}}$, or $\eta_{\mathrm{sw}}$ and $\eta_{\mathrm{rt}}$ within the same plot. \textbf{(e-h)} Similar to (a-d) except for that the magnetic field $B_{\mathrm{x}}^{'}$ applied in parallel to the current $I_{\mathrm{b}}$. All data here are averaged from $N=30$ measurements at $V_{\mathrm{bg}}=0$ and base temperature. The current ramp rate is $\sim$9\,nA/s.} 
\label{figure:2}
\end{figure}

 Figure \ref{figure:2} shows the magnetic field dependence of the JDE along two different directions, $B_{\mathrm{x}}^{'}$ and $B_{\mathrm{y}}^{'}$. Here, $B_{\mathrm{x}}^{'}$ and $B_{\mathrm{y}}^{'}$ represent configurations where the in-plane magnetic field $B_{\mathrm{xy}}$ is parallel and perpendicular to the direction of $I_{\mathrm{b}}$, respectively (see Figure \ref{figure:1}b). Figure \ref{figure:2}a presents the absolute values of $I_{\mathrm{sw}}^{+}$ and $I_{\mathrm{sw}}^{-}$  as a function of $B_{\mathrm{y}}^{'}$ and Figure \ref{figure:2}b shows the results for retrapping currents. In the two figures, both switching and retrapping currents decrease with increasing magnetic field, caused by the suppression of induced superconductivity. Meanwhile, a nonzero difference between $I_{\mathrm{sw}}^{+}$ and $|I_{\mathrm{sw}}^{-}|$ (as well as between $I_{\mathrm{rt}}^{+}$ and $|I_{\mathrm{rt}}^{-}|$) can be recognized. Figure \ref{figure:2}c shows $-\Delta I_{\mathrm{sw}}$ and $\Delta I_{\mathrm{rt}}$ as a function of $B_{\mathrm{y}}^{'}$(solid dots). In the figure, both $\Delta I_{\mathrm{sw}}$ (in pink) and $\Delta I_{\mathrm{rt}}$ (in blue) exhibit antisymmetric behaviors with respect to $B_{\mathrm{y}}^{'}$, i.e., $\Delta I_{\mathrm{sw}}(B_{\mathrm{y}}^{'})=-\Delta I_{\mathrm{sw}}(-B_{\mathrm{y}}^{'})$ and $\Delta I_{\mathrm{rt}}(B_{\mathrm{y}}^{'})=-\Delta I_{\mathrm{rt}}(-B_{\mathrm{y}}^{'})$, with  maximum and minimum values occurring at $B_{\mathrm{y}}^{'}\sim\,\pm 50\,\rm{mT}$. Figure \ref{figure:2}d exhibits diode efficiency $-\eta_{\mathrm{sw}}$ and $\eta_{\mathrm{rt}}$ as a function of $B_{\mathrm{y}}^{'}$, and the maximum values of efficiency exceed $2\%$ for both $\eta_{\mathrm{sw}}$ and $\eta_{\mathrm{rt}}$. In Figures \ref{figure:2}a-\ref{figure:2}d, $B_{\mathrm{y}}^{'}$ is swept from negative to positive, and similar results can be observed when the field sweep direction is reversed (i.e., from positive to negative, see Figure S4 in the Supporting Information). When the in-plane magnetic field is parallel to bias current, suppression of switching current and retrapping current could also be observed, as shown in Figure \ref{figure:2}e and \ref{figure:2}f. However, almost no signal of the JDE is observed, as can be seen in Figure \ref{figure:2}g and \ref{figure:2}h. The JDE measurements as in Figure 2 have been done on device B and similar results are observed (see details in Figure S6 in the Supporting Information). In Figure \ref{figure:2}, both $\Delta I_{\mathrm{sw}}$ and $\Delta I_{\mathrm{rt}}$ are nearly zero throughout the measurement range of $B_{\mathrm{x}}^{'}$. This excludes a possible explanation that $B_{\mathrm{x}}^{'}$ or $B_{\mathrm{y}}^{'}$ has a residual out-of-plane component leading to the JDE.\cite{hou2023ubiquitous} Based on our previous work, the Rashba-type SOI is dominant in InAs nanosheets, with its direction perpendicular to the bias current.\cite{fan2022electrically} Hereafter, the results are analyzed in the context of finite momentum pairing induced by Rashba SOI and magnetic field. When a magnetic field is applied parallel to the SOI field (i.e., $B_{\mathrm{y}}^{'}$ in our device), a momentum shift $q_{\mathrm{0}}$ is acquired for Cooper pairs when crossing the Josephson junction.\cite{davydova2022universal,zhang2022general} Consequently, a phase shift $\delta =2q_{\mathrm{0}}d$ is present for supercurrent across the junction,\cite{yokoyama2014anomalous,szombati2016josephson} where $d$ is the gap length between the two superconducting leads. At small magnetic fields, $ q_{\mathrm{0}}$ is linear in ${B_{\mathrm{y}}}^{'}$, so $\delta\approx\pi\frac{B_{\mathrm{y}}^{'}} {B_{\mathrm{d}}}$, where ${B_{\mathrm{d}}}$ is a parameter dependent on the junction's geometry and material properties. The finite phase shift together with high-quality semiconductor-superconductor interface (see Figures S2 and S3 in the Supporting Information) 
 would ultimately result in the JDE.\cite{baumgartner2022supercurrent,davydova2022universal,pal2022josephson,zhang2022evidence,mazur2022gate} Following Ref.\zcite{pal2022josephson}, the switching current difference  is governed by the equation: \begin{equation*} \Delta I_{\mathrm{sw}} \propto \left[1-\left(\frac{|B|}{B_{\mathrm{c}}}\right)^{2}\right]^{2} \sin \left(\pi\frac{B_{\mathrm{y}}^{'}}{B_{\mathrm{d}}} \right)\end{equation*}where $B$ is the applied magnetic field and ${B_{\mathrm{c}}}$ is the critical magnetic field of the Josephson junction.\cite{davydova2022universal,pal2022josephson,mazur2022gate} In Figure \ref{figure:2}c, $\Delta I_{\mathrm{sw}}$ and  $\Delta I_{\mathrm{rt}}$ are fitted with the above formula, where $B_{\mathrm{c}}$ is taken as $B_{\mathrm{c}}\sim\,270\,\rm mT$ (see Figure S1 in the Supporting Information), and the fitting curves are displayed as pink and blue solid lines. The well agreement between the data and the fitting curves is consistent with the scenario that finite-momentum pairing gives rise to the JDE observed in our devices.
 \par 
When an external magnetic field is applied, both self-field and orbital effects can, in principle, also induce a JDE. However, these two mechanisms can be excluded as the origin of the JDE observed in our devices. Although the self-field effect arising from non-straight current flow may lead to a JDE, it occurs only under an out-of-plane magnetic field.\cite{golod2022demonstration} Therefore, it cannot account for the JDE observed under in-plane fields. The orbital effect under an in-plane magnetic field could also induce a JDE, but it typically requires a finite cross-sectional area to form a loop, as discussed in Ref.~\zcite{banerjee2023phase}. In our devices, however, the Al electrodes are in direct contact with the InAs nanosheet, making the formation of such a loop unlikely. Consequently, it is difficult for a sizable loop to develop under in-plane magnetic fields of several tens of millitesla, ruling out the orbital effect as the mechanism responsible for the observed JDE in our devices.

\par
\begin{figure}[bt]
    \centering
    \includegraphics[width=0.8\textwidth]{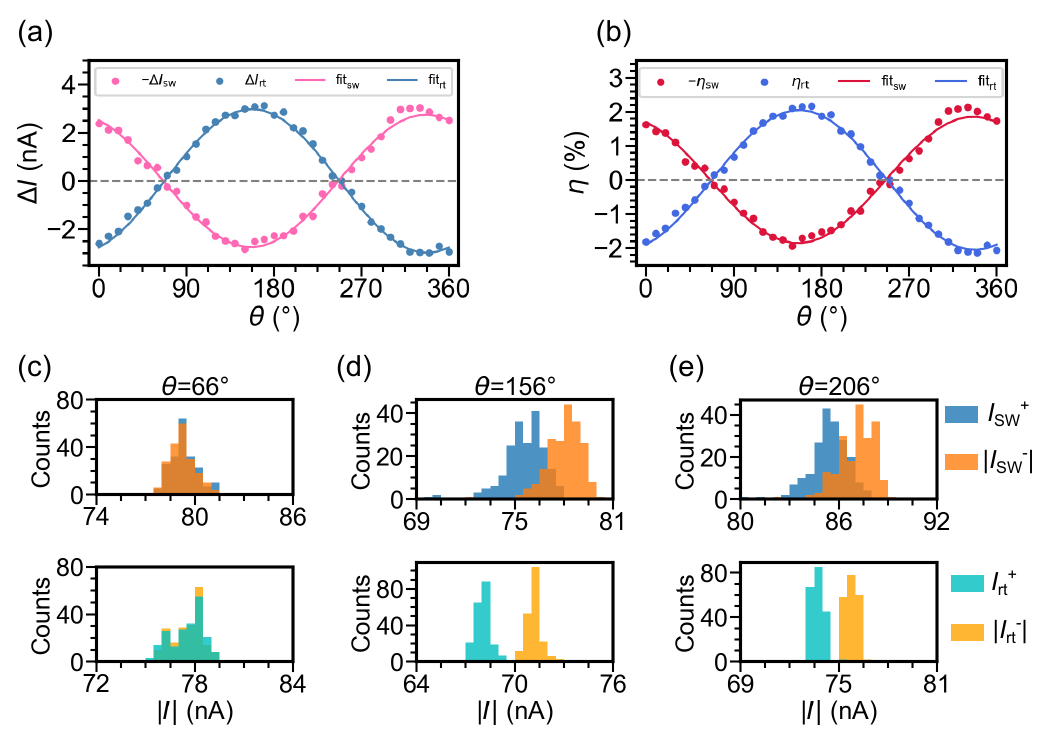} 
    \caption{
    \textbf{Dependence of the JDE on in-plane magnetic field orientation.}
    \textbf{(a)} Switching current difference (\(-\Delta I_{\mathrm{sw}}\), pink dots) and retrapping current difference (\(\Delta I_{\mathrm{rt}}\), blue dots) as a function of the angle \(\theta\) between the magnetic field \(B_{xy}\) and the \(x\)-axis. Sinusoidal fits capture the dependence on \(\theta\). 
    \textbf{(b)} Diode efficiencies (\(-\eta_{\mathrm{sw}}\) and \(\eta_{\mathrm{rt}}\)) as a function of \(\theta\). 
    \textbf{(c-e)} Histograms of switching currents(upper panels) and retrapping currents(lower panels)  for 200 measurements at in-plane magnetic field orientations \(\theta = 66^\circ\) (\(B_{xy} \parallel I_{\mathrm{b}}\)), \(\theta = 156^\circ\) (\(B_{xy} \perp I_{\mathrm{b}}\)), and \(\theta = 206^\circ\). All measurements were performed at \(|B_{xy}| = 50\,\text{mT}\), \(V_{\mathrm{bg}} = 0\) and base temperature.The current ramp rate is $\sim$9\,nA/s.
    }
    \label{figure:3}
\end{figure}

 Figure \ref{figure:3} displays the dependence of the JDE on the direction of in-plane magnetic field with a fixed magnitude $|B_{\mathrm{xy}}|=50\,\rm mT$. In Figure \ref{figure:3}a, $-\Delta I_{\mathrm{sw}}$ (pink dots) and $\Delta I_{\mathrm{rt}}$ (blue dots) are plotted as a function of $\theta$, where $\theta$ represents the angle between $B_{\mathrm{xy}}$ and $B_{\mathrm{x}}$ (see Figure \ref{figure:1}b). Figure \ref{figure:3}b shows the dependence of diode efficiency $-\eta_{\mathrm{sw}}$ (red dots) and $\eta_{\mathrm{rt}}$ (blue dots) on angle $\theta$. In the two figures, the gray dashed lines indicate the positions where $\Delta I=0$ and $\eta =0$, respectively, and the solid lines are sinusoidal fits to  the data points. From the sinusoidal fits, the angles at which the JDE is strongest are found to be $\theta\sim 156\,\degree$ and $336\,\degree$, while the angles at which the JDE become weakest are found to be $\theta\sim 66\,\degree$ and $246\,\degree$. Note that at $\theta = 156\,\degree$, $B_{\mathrm{xy}}$ is aligned with $B_{\mathrm{y}}^{'}$. Figures \ref{figure:3}c-\ref{figure:3}e show histogram distributions from $N=200$ measurements for switching current (upper panels) and retrapping current (lower panels) at $\theta$ equals to $66\,\degree$ ($B_{\mathrm{xy}}\parallel I_{\mathrm{b}}$), $156\,\degree$ ($B_{\mathrm{xy}}\perp I_{\mathrm{b}}$) and $206\,\degree$ (in between), respectively. These distributions reveal a clear trend of a more pronounced JDE as $B_{\mathrm{xy}}$ rotates from being parallel to being perpendicular to $I_{\mathrm{b}}$. As discussed above, finite momentum pairing takes place when magnetic field $B_{\mathrm{xy}}$ and SOI field $B_{\mathrm{SOI}}$ are in parallel ($B_{\mathrm{xy}} \parallel B_{\mathrm{SOI}}$) and, thus, a finite phase shift emerges for Cooper pairs transporting along a perpendicular direction (i.e., $I_{\mathrm{b}} \perp B_{\mathrm{xy}}$ or $I_{\mathrm{b}} \perp B_{\mathrm{SOI}}$).\cite{zhang2022general}  When $B_{\mathrm{xy}}$ is rotated from being parallel to being perpendicular to $I_{\mathrm{b}}$, the component of the field along $B_{\mathrm{SOI}}$ increases, therefore enhancing the finite momentum pairing and the JDE. This is validated by the sinusoidal shape of the results in Figures \ref{figure:3} a and \ref{figure:3} b. In a nutshell, the angle-dependence of the JDE shows well agreement with the mechanism of finite momentum pairing induced by Rashba SOI and magnetic field.

\subsection{Gate Voltage and Temperature Dependence}

\begin{figure}[t]
    \centering
    \includegraphics[width=0.8\textwidth]{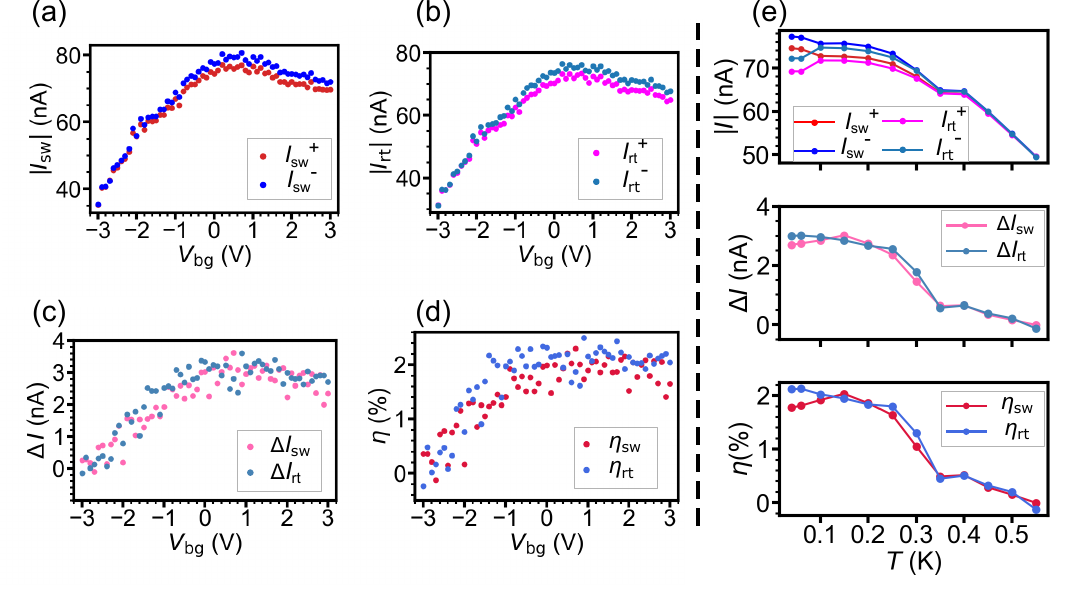} 
    \caption{
    \textbf{Dependence of the JDE on gate voltage and temperature.}
    \textbf{(a)} and \textbf{(b)} Absolute values of switching currents ($I_{\mathrm{sw}}^+$ and $|I_{\mathrm{sw}}^-|$) and retrapping currents ($I_{\mathrm{rt}}^+$ and $|I_{\mathrm{rt}}^-|$) as a function of back gate voltage \(V_{\mathrm{bg}}\). \textbf{(c)} and \textbf{(d)} Absolute values of current differences ($|\Delta I_{\mathrm{sw}}|$ and $\Delta I_{\mathrm{rt}}$) and diode efficiencies ($|\eta_{\mathrm{sw}}|$ and \(\eta_{\mathrm{rt}}\)) as a function of \(V_{\mathrm{bg}}\). The results show a monotonic suppression of the JDE as \(V_{\mathrm{bg}}\) decreases from \(0.5\,\text{V}\) to \(-3\,\text{V}\), with a complete suppression occurring at \(V_{\mathrm{bg}} \sim -3\,\text{V}\). The data in (a)-(d) were taken at \(B'_y = 50\,\text{mT}\) and base temperature. The current ramp rate is $\sim$33\,nA/s. \textbf{(e)} Temperature dependence of the JDE as shown by switching and retrapping currents (upper panel), supercurrent differences (middle panel), and diode efficiencies (lower panel). Measurements were conducted at \(B'_y = 50\,\text{mT}\) and \(V_{\mathrm{bg}} = 0\,\text{V}\). The current ramp rate is $\sim$9\,nA/s.
}
    \label{figure:4}
\end{figure}
\par In principle, finite momentum pairing in superconductor-semiconductor hybrid Josephson junctions can originate from screening effect or interaction between Zeeman effect and SOI.\cite{davydova2022universal,yuan2022supercurrent,pal2022josephson,banerjee2023phase} Studying the dependence of the JDE on gate voltage is helpful for tracking its microscopic origin.\cite{davydova2022universal,yuan2022supercurrent} Figure \ref{figure:4}a shows switching current, $I_{\mathrm{sw}}^{+}$ and |$I_{\mathrm{sw}}^{-}$|, as a function of back gate voltage $V_{\mathrm{bg}}$ measured for device A at \(B'_y = 50\,\text{mT}\). The supercurrent keeps increasing for $V_{\mathrm{bg}}$ from \(-3\) to \(0.5\,\mathrm{V}\) and drops slightly from \(0.5\) to \(3\,\mathrm{V}\). Similar behaviors are observed for retrapping current as shown in Figure \ref{figure:4}b. The supercurrent rise in the gate range of \(-3\) to \(0.5\,\mathrm{V}\) is caused by the increase of carrier density, which is supported by the decrease of the normal resistance of the device (see Figure S2 in the Supporting Information). The slight drop in the supercurrent when the back-gate voltage increases from  $0.5\,\mathrm{V}$ to more positive values also occurs without an external magnetic field (see Figure S2b in the Supporting Information), which could be attributed to the change in junction transparency, or gate-controlled induced superconductivity,\cite{van2023electrostatic} or gate-controlled interference between transverse subbands.\cite{sriram2019supercurrent} Figures \ref{figure:4}c and \ref{figure:4}d display the gate dependence of the supercurrent differences ($\Delta I_{\mathrm{sw}}$ and $\Delta I_{\mathrm{rt}}$) and the diode efficiencies ($\eta_{\mathrm{sw}}$ and $\eta_{\mathrm{rt}}$). The JDE shows a monotonic suppression as $V_{\mathrm{bg}}$ decreases from 0.5 to $ -3\rm \,V$. Notably, at $V_{\mathrm{bg}}\sim \,-3\rm \, V$, the JDE completely vanishes, with both $\Delta I_{\mathrm{sw}}\sim \,0$ and $\Delta I_{\mathrm{rt}}\sim \,0$. In contrast, a significant amount of supercurrent is still conserved at $V_{\mathrm{bg}}=-3\rm \,V$ as seen in Figures \ref{figure:4}a and \ref{figure:4}b. Such a gate-controlled JDE indicates that screening effect/diamagnetic effect is not the origin of the non-reciprocal supercurrent in this work. This is because the JDE induced by screening effect/diamagnetic effect relies on the diamagnetic supercurrent in the superconducting regions and should not be turned off by gate voltage.\cite{yuan2022supercurrent,davydova2022universal,sundaresh2023diamagnetic} The suppression of the JDE at more negative gate voltages, as seen in Figures 4c and 4d, could be caused by the change of junction transparency or the suppression of finite momentum pairing. As indicated in Figure S2 of the Supporting Information, the interface transparency increases with more negative gate voltages. While the influence of transparency on Josephson diode effect is complex, merely increasing the transparency cannot lead to a monotonic decrease and eventual complete suppression of the JDE.\cite{costa2023microscopic} According to our previous studies, Rashba SOI can exist even at zero gate voltage due to band bending effect induced by the asymmetric dielectric environments of the nanosheets, and the Rashba SOI can be quenched at a gate voltage that compensates for the band bending.\cite{fan2022electrically,chen2021strong} Therefore, the back-gate–tuned suppression and eventual disappearance of the JDE observed in Figures \ref{figure:4}c and \ref{figure:4}d is likely caused by the suppression of finite-momentum pairing via gate-controlled SOI. When $V_{\mathrm{bg}}$ increases from 0.5 to $3\rm \,V$, the JDE displays behavior characterized by near-saturation or a slight decline, as seen in Figures \ref{figure:4}c and \ref{figure:4}d. This is likely due to a competing effect among tunable parameters such as chemical potential, SOI strength, and interface transparency. Note that the gate voltage in all measurements was limited to $\pm$3 V to avoid potential dielectric breakdown and the device behavior beyond this range remains unexplored. Figure \ref{figure:4}e shows temperature dependence of the JDE. As the temperature increases, both the switching and retrapping currents exhibit a nearly monotonic decrease, which is attributed to the closing of the superconducting gap. Around \(T \sim 550 \, \mathrm{mK}\), the remaining supercurrent is approximately 50\,nA. As a comparison, both the supercurrent differences, $\Delta I_{\mathrm{sw}}$ and $\Delta I_{\mathrm{rt}}$, and the diode efficiencies, $\eta_{\mathrm{sw}}$ and $\eta_{\mathrm{rt}}$, drop to zero as $T$ increases to \(550 \, \mathrm{mK}\). The complete suppression of the JDE with increasing temperature has been observed in previous studies and has been attributed to a temperature-induced weakening of higher harmonics in the current–phase relation.\cite{turini2022josephson,pal2022josephson}

\section{Conclusion} 
\par In conclusion, we have demonstrated the existence of intrinsic Josephson diode effect (JDE) in planar Josephson junctions made from InAs nanosheets with superconducting aluminum contacts. The JDE is likely driven by finite-momentum pairing, which arises from the interplay between Rashba SOI and Zeeman splitting under an in-plane magnetic field. In experiments, we see a strong dependence of the JDE on magnetic field orientation and the JDE is maximized when the magnetic field is aligned with the expected spin-orbit field. Importantly, we observe that the JDE is highly tunable via electrostatic gating with a complete suppression at specific gate voltages, indicating the critical role of the Rashba SOI in enabling the non-reciprocal superconducting current. Our findings contribute to the growing body of research demonstrating that JDE is not only a fundamental phenomenon to explore novel physics of nonreciprocal superconductivity but also holds promise for applications in superconducting electronics, spintronics, and quantum information technology. The ability to continuously control JDE through gate voltage offers exciting prospects for developing low-power, non-dissipative components for superconducting circuits, including rectifiers, memory elements, and logic devices.

\section*{Author contributions}
H.Q.X conceived and supervised the project. S.Y., H.S., H.G and X.W. fabricated the devices. D.P. and J.Z. grew the InAs nanosheets. S.Y., J.Y.W.  and Y.L. performed the transport measurements. S.Y., J.Y.W. and H.Q.X. analyzed the measurement data. S.Y., J.Y.W. and H.Q.X. wrote the manuscript with inputs from all the authors.  

\section*{Acknowledgements}
The authors would like to thank Dr. Po Zhang for his assistance with the measurement codes. This work is supported by the National Natural Science Foundation of China (Grant Nos.
 92165208, 12374480, 11874071 and 12374459). D.P. acknowledges the support from Youth Innovation Promotion Association, Chinese Academy of Sciences (Nos. 2017156 and Y2021043).
 
\section*{Data analysis and data availability} 
The raw data and the analysis files are available upon request.

\section*{conflict of interest}
The authors declare no conflict of interests. 

\section*{Supporting Information}
The Supporting Information includes: details of device fabrication and additional transport experiment data (PDF).

\bibliography{references}

\end{document}


\newpage
\section{Methods}

\subsection*{Device fabrication} 

 \par Back-gate electrodes were fabricated using electron-beam lithography followed by deposition of 5/25\, nm Ti/Au via electron-beam evaporation on a SiO$_2$/Si substrate. Subsequently, a 15\,nm dielectric layer of $\mathrm{HfO_2}$ was grown by atomic layer deposition (ALD). InAs nanosheets with typical thicknesses of 15–30\,nm were then mechanically transferred onto the pre-patterned back gates. Electron-beam lithography was employed again to define the contact areas. After development at room temperature, the native oxide on the contact areas of the InAs nanosheets was removed using a diluted $\mathrm{(NH_4)_2S_x}$ solution. Immediately afterward, a 5\,nm Ti layer was deposited via electron-beam evaporation, followed by a deposition of a 50\,nm Al layer using thermal evaporation. A 15\,nm $\mathrm{Al_2O_3}$ layer was then deposited via ALD, and finally, a 5/50\,nm Ti/Au layer was deposited using electron-beam evaporation to define the top-gate electrodes. Devices were pre-selected using a room-temperature probe station for subsequent low-temperature transport measurements.
\newpage
\subsection*{Low temperature transport measurements} 
 \par Low-temperature measurements were carried out in a dilution refrigerator equipped with a vector magnet. The devices were measured with a typical quasi-four-terminal measurement setup, in which bias current $I_{\mathrm{b}}$ is applied to superconducting leads while voltage drop across the junctions $V$ is measured. Since the switching and retrapping currents in a Josephson junction are typically stochastic, repeated measurements are beneficial for accurate analysis of Josephson diode effect (JDE). Therefore, at each parameter setting (such as magnetic field and gate voltage), $V-I_{\mathrm{b}}$ traces were taken for many times. Then averaged switching current and averaged retrapping current are used to quantify the JDE. The top gates of both devices in this work showed negligible tunability, likely due to the screening effect of contact electrodes. Thus, the top gates were set to zero throughout the measurements.

\newpage
 \section*{Additional Data}
\subsection*{Additional data of device A}
\begin{figure*}[h!]
\centering
\includegraphics[width=0.9\linewidth]{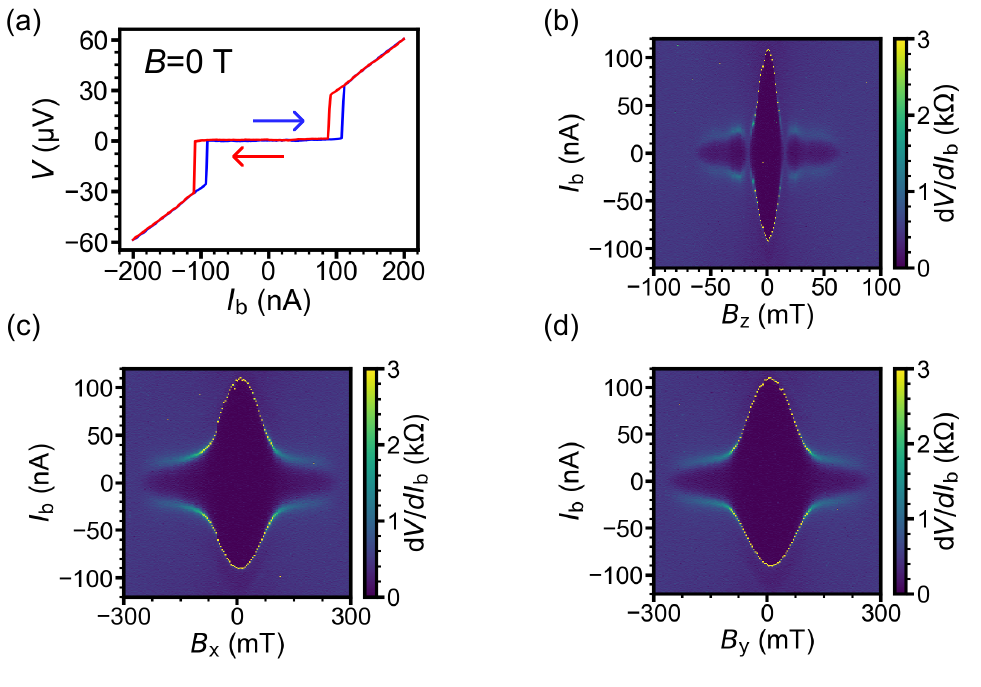} 
\caption{
\doublespacing \textbf{Basic superconducting properties of device A.} \textbf {(a)} Voltage across the junction $V$ as a function of bias current $I_{\mathrm{b}}$ at zero magnetic field. Blue and red traces represent the results of upward and downward current sweeps, respectively.
\textbf{(b)} Differential resistance $dV/dI_{\mathrm{b}}$ as a function of $I_{\mathrm{b}}$ and out-of-plane magnetic field $B_{\mathrm{z}}$. The central dark area is the superconducting region with $dV/I_{\mathrm{b}} =0$. \textbf{(c)} Differential resistance $dV/dI_{\mathrm{b}}$ as a function of $I_{\mathrm{b}}$ and in-plane magnetic field $B_{\mathrm{x}}$. \textbf{(d)} Differential resistance $dV/dI_{\mathrm{b}}$ as a function of $I_{\mathrm{b}}$ and in-plane magnetic field $B_{\mathrm{y}}$. All data were taken at $V_\mathrm{bg}=0$ and base temperature.
}\label{figure:S1}
\end{figure*}

\begin{figure*}[h!]
\centering
\includegraphics[width=0.9\linewidth]{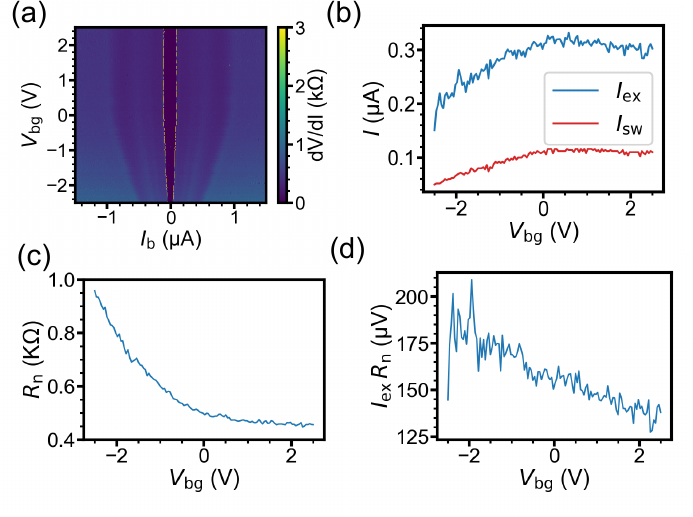} 
\caption{
\doublespacing \textbf{Gate-dependent supercurrent measurements of Device A} 
\textbf {(a)} $dV/dI_{\mathrm{b}}$ versus  $I_{\mathrm{b}}$ and back gate voltage $V_{\mathrm{bg}}$. \textbf{(b)} Switching current $I_{\mathrm{sw}}$ and excess current $I_{\mathrm{ex}}$ extracted from (a) as a function of $V_{\mathrm{bg}}$. \textbf{(c)} Normal-state resistance $R_{\mathrm{n}}$ extracted from (a) as a function of $V_{\mathrm{bg}}$. \textbf{(d)} Product $I_{\mathrm{ex}}R_{\mathrm{n}}$ as a function of $V_\mathrm{bg}$. The value of $I_\mathrm{ex}R_{\mathrm{n}}$ varies between 130 and $200\,\mathrm{\mu V}$, leading to an interface transparency ranging from $72\,\%$ to $85\,\%$. All data were measured at $B=0$ and base temperature.
}\label{figure:S2}
\end{figure*}

\begin{figure*}[h!]
\centering
\includegraphics[width=0.9\linewidth]{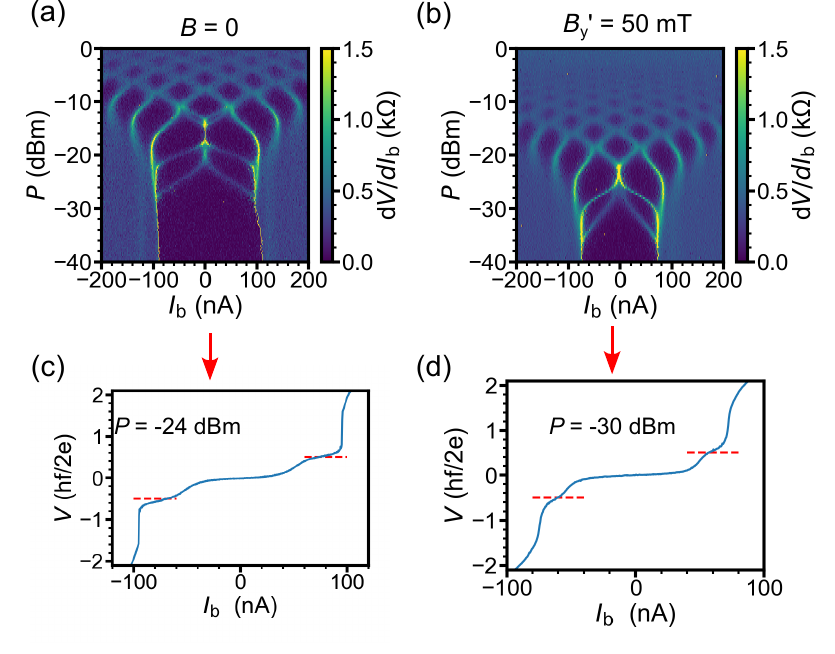} 
\caption{
\doublespacing \textbf{Half-integer Shapiro steps observed in device A.} 
\textbf {(a)} Differential resistance $dV/I_{\mathrm{b}}$ as a function of bias current $I_{\mathrm{b}}$ and microwave power at $f=6.6 \,\mathrm{GHz}$ and $B=0$. \textbf{(b)} Differential resistance $dV$/$dI_{\mathrm{b}}$ as a function of $I_{\mathrm{b}}$ and microwave power at $f=7 \,\mathrm{GHz}$ and $B_{\mathrm{y}}^{'}=50\,\mathrm{mT}$. \textbf{(c)} and \textbf{(d)} Line cuts from (a) and (b), respectively, showing the presence of half-integer Shapiro steps. The appearance of these half-integer Shapiro steps implies the presence of higher harmonics in the current-phase relation in the device.
}
\label{figure:S3}
\end{figure*}

\begin{figure*}[h!]
\centering
\includegraphics[width=0.9\linewidth]
{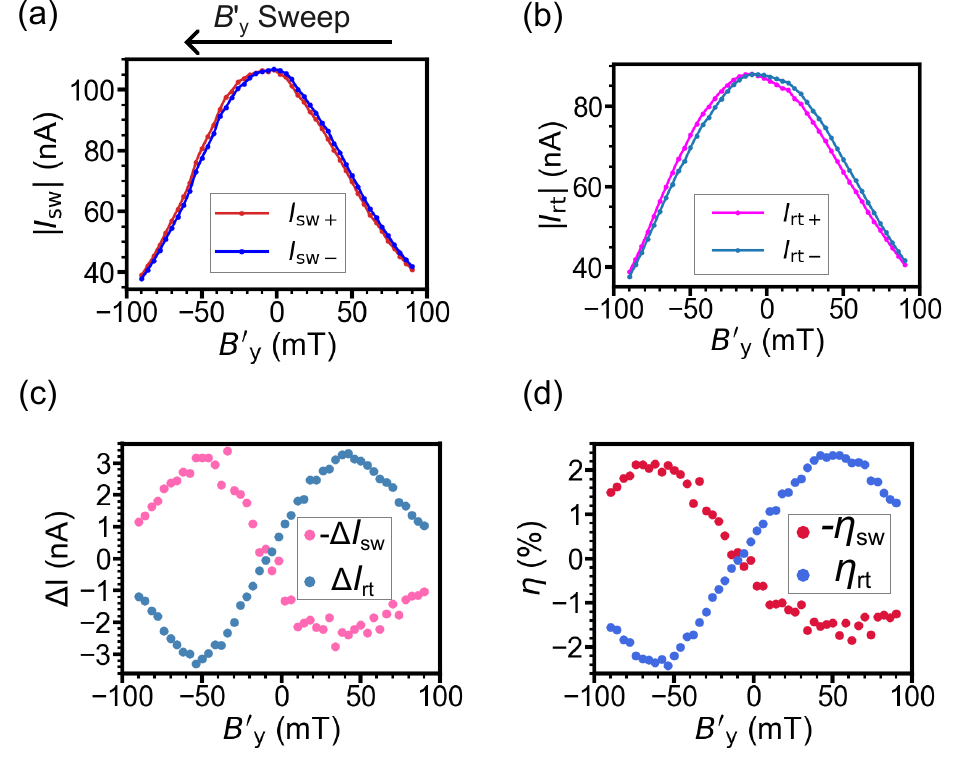}
\caption{
\doublespacing \textbf{Dependence of the JDE on in-plane magnetic field $B_{\mathrm{y}}^{'}$ swept from positive to negative values.} \textbf{(a)} Switching current $I_{\mathrm{sw}}^{+}$ and $|I_{\mathrm{sw}}^{-}|$ as a function of $B_{\mathrm{y}}^{'}$. \textbf{(b)} Retrapping current $I_{\mathrm{rt}}^{+}$ and $|I_{\mathrm{rt}}^{-}|$ as a function of  $B_{\mathrm{y}}^{'}$. \textbf{(c)} Pink: difference between the absolute values of the switching currents for forward and backward sweeps ($\Delta I_{\mathrm{sw}}=I_{\mathrm{sw}}^{+}-|I_{\mathrm{sw}}^{-}|$), extracted from (a), as a function of $B_{\mathrm{y}}^{'}$. Blue: difference between the absolute values of the retrapping currents for forward and backward sweeps ($\Delta I_{\mathrm{rt}}=|I_{\mathrm{rt}}^{-}|-I_{\mathrm{rt}}^{+}$), extracted from (b), as a function of $B_{\mathrm{y}}^{'}$. \textbf{(d)} Diode efficiency $-\eta_{\mathrm{sw}}$=($I_{\mathrm{sw}}^{+}$-|$I_{\mathrm{sw}}^{-}$|)/($I_{\mathrm{sw}}^{+}$+|$I_{\mathrm{sw}}^{-}|$) and $\eta_{\mathrm{rt}}$=(|$I_{\mathrm{rt}}^{-}$|-$I_{\mathrm{rt}}^{+}$)/($I_{\mathrm{rt}}^{+}$+|$I_{\mathrm{rt}}^{-}|$) as a function of $B_{\mathrm{y}}^{'}$. All data were taken at $V_{\mathrm{bg}}=0$ and base temperature, and were averaged from $N=30$ measurements. Here, $B_{\mathrm{y}}^{'}$ is stepped downward in comparison with Figure 2 in the main article where $B_{\mathrm{y}}^{'}$ is stepped upward.
}
\label{figure:S4}
\end{figure*}

\clearpage

\subsection*{Additional data of device B.} 
\begin{figure*}[h]
\centering
\includegraphics[width=0.9\linewidth]{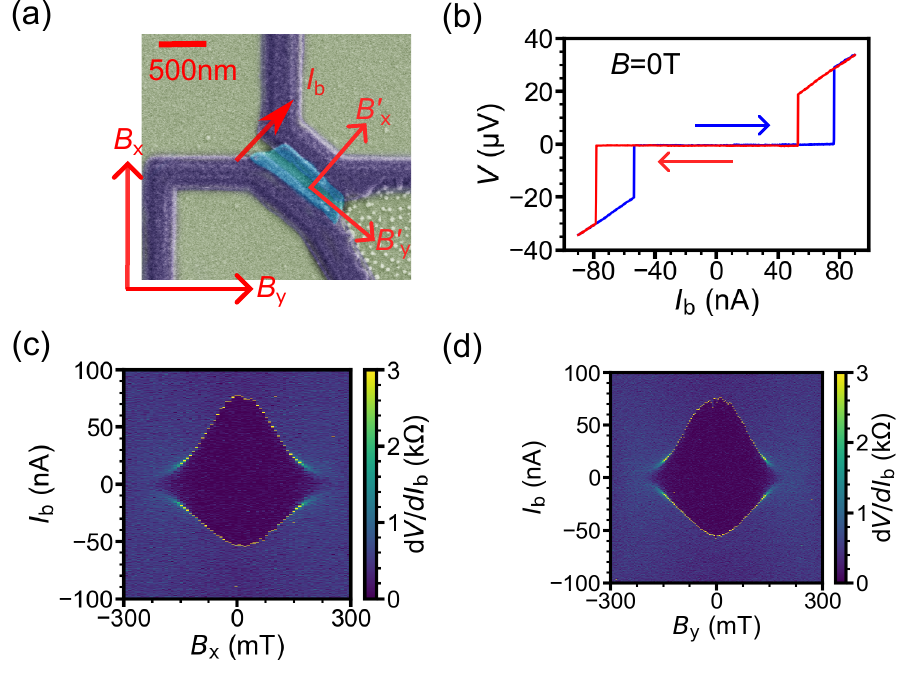}
\caption{
\doublespacing \textbf{Basic properties of device B.} \textbf{(a)} False-colored SEM image of device B, together with magnetic field directions and bias current direction. \textbf{(b)} Typical hysteresis $V-I_{\mathrm{b}}$ curves of device B. \textbf{(c)} Differential resistance $dV$/$dI_{\mathrm{b}}$ as a function of $I_{\mathrm{b}}$ and in-plane magnetic field $B_{\mathrm{x}}$. \textbf{(d)} Differential resistance $dV/dI_{\mathrm{b}}$ as a function of $I_{\mathrm{b}}$ and in-plane magnetic field $B_{\mathrm{y}}$. The data in (b-d) were taken at $V_\mathrm{bg}=0$ and base temperature.
}
\label{figure:S5}
\end{figure*}

\begin{figure*}[h!]
\centering
\includegraphics[width=1\linewidth]{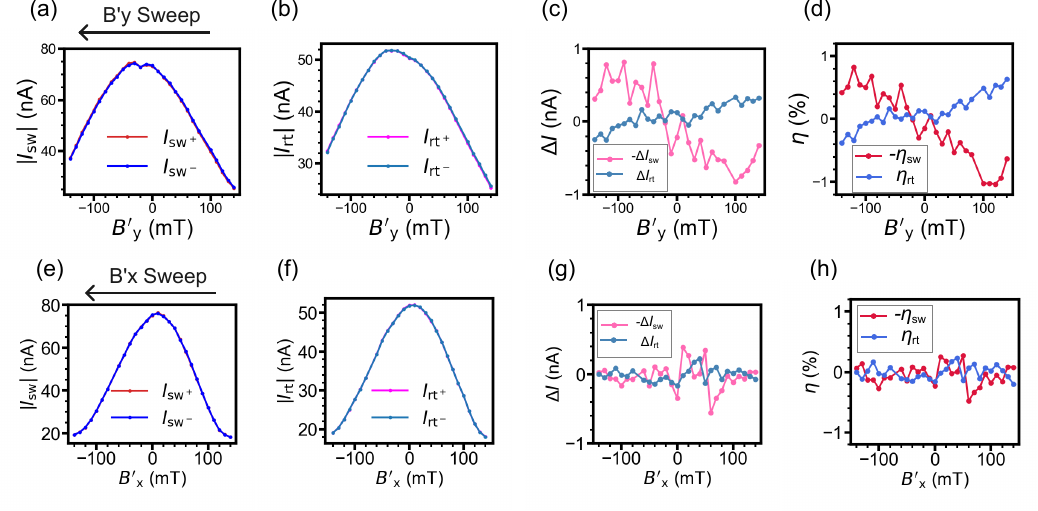}
\caption{\doublespacing \textbf{The JDE in device B.} \textbf{(a-d)} Diode effect observed when the in-plane magnetic field is perpendicular to bias current, i.e. parallel to the effective Rashba field ($B_\mathrm{SOI}$). \textbf{(e-h)} Negligible diode effect observed when the in-plane magnetic field is perpendicular to $B_\mathrm{SOI}$.  
}\label{figure:S6}
\end{figure*}